\documentclass[superscriptaddress,nofootinbib,notitlepage,10pt,amsmath,amssymb,aps,pra]{revtex4-1}

\usepackage[a4paper, margin=2.50cm]{geometry}

\usepackage{bbm}
\usepackage{graphicx}
\usepackage{fancyhdr, color}
\usepackage{hyperref}

\begin{document}

%\fancyhead[R]{\sc \color[rgb]{0.4,0.2,0.9}{Quantum Thermodynamics book}}
\fancyhead[R]{}

\title{Nonequilibrium many-body quantum dynamics:\texorpdfstring{\\}{} from full random matrices to real systems}

\author{Lea F. Santos}
\email{lsantos2@yu.edu} 
\affiliation{Department of Physics, Yeshiva University, New York, New York 10016, USA.}

\author{E. J. Torres-Herrera}
\email{etorresh@ifuap.buap.mx} 
\affiliation{Instituto de F{\'i}sica, Benem\'erita Universidad Aut\'onoma de Puebla, Apartado Postal J-48, Puebla, Puebla, 72570, M\'exico}

\date{\today}

\begin{abstract}

We present an overview of our studies on the nonequilibrium dynamics of quantum systems that have many interacting particles. Our emphasis is on systems that show strong level repulsion, referred to as chaotic systems. We discuss how full random matrices can guide and support our studies of realistic systems. We show that features of the dynamics can be anticipated from a detailed analysis of the spectrum and  the structure of the initial state projected onto the energy eigenbasis. On the other way round, if we only have access to the dynamics, we can use it to infer the properties of the spectrum of the system. Our focus is on the survival probability, but results for other observables, such as the spin density imbalance and Shannon entropy are also mentioned.

\end{abstract}

\maketitle

\thispagestyle{fancy}

%% Introduction %%%%%%%%%%%%
\section{Introduction} 
\label{sec:introduction}

More than 90 years after the derivation of the Schr\"odinger equation, studies of the evolution of isolated finite quantum systems still receive much theoretical and experimental attention. This may be surprising, since we are simply dealing with linear dynamics. However, the complexity of some of these quantum systems lies in the effects of the interactions between their many particles, which in most cases prevent analytical results and limits numerical studies to small system sizes. Several open questions permeate the field, including the description of the relaxation to equilibrium and the viability of thermalization~\cite{Borgonovi2016}, the existence of a localized phase  in the presence of disorder~\cite{LuitzAP}, the possibility of saturating bounds for the quantum speed limit~\cite{GiovannettiPRA2003}, and the quantum-classical correspondence~\cite{Akila2017}. 

For one-body chaos, the quantum-classical correspondence is well understood. There are conjectures~\cite{Bohigas1984} and numerical studies~\cite{Casati1980} connecting classical chaos with specific properties of the spectrum of quantum systems and also semiclassical methods that bridge the classical and quantum domains~\cite{Gutzwiller1971,Strutinsky1976}. Specifically, quantum systems whose classical counterparts are chaotic have correlated eigenvalues which repel each other resulting in a rigid spectrum. This is to be contrasted with regular (integrable) systems, where the eigenvalues are uncorrelated. In chaotic systems, the distribution of the spacings of neighboring levels follows the Wigner-Dyson distribution~\cite{MehtaBook,Guhr1998} instead of the Poissonian distribution that is usually found in regular systems. 

When it comes to interacting many-body quantum systems, the quantum-classical correspondence is still missing. This is ironic, because quantum chaos studies have in fact been strongly stimulated from Wigner's studies of many-body nuclear systems~\cite{Wigner1951P,Wigner1958}. One often extends the findings from one-body chaos and refers to many-body quantum systems with Wigner-Dyson distributions as chaotic systems, as we do in this chapter, but this link has not been proved yet.  The subject has recently received much attention~\cite{ScaffidiARXIV,SchmittARXIV,BorgonoviARXIV}, leading to significant progress in the development of semiclassical analysis for interacting many-body systems~\cite{Akila2017,DoronCohen}. 

Along the years, there have also been several attempts to connect exponential behaviors in the quantum domain with classical chaos. This topic has been once again revived now in the context of the so-called out-of-time-ordered four-point correlator (OTOC). While exponential behaviors for the OTOC have been confirmed in some cases of one-body chaos~\cite{Rozenbaum2017}, but not all \cite{Hashimoto2017}, there is no consensus on what we might expect for many-body systems~\cite{Luitz2017}.

In a new study~\cite{BorgonoviARXIV}, it has been shown that the  number of states participating in the quantum evolution of realistic interacting many-body quantum systems perturbed far from equilibrium grows exponentially fast in time. This number, known as participation ratio or number of principal components,  is an example of an OTOC. The results are numerical and semi-analytical. The quantum-classical correspondence in this work comes from the analogy between the growth of the volume filled by the states participating in the evolution  and the growth of the volume of the classical phase-space visited by the classical trajectories. For both, the growth is exponential and the rate should be given by the Kolmogorov-Sinai entropy.

In this chapter, we focus on the survival probability, which is a two-point correlation function, and study its entire evolution after a quench. We show that the dynamics at short times cannot distinguish between realistic many-body quantum systems with and without level repulsion~\cite{Torres2014PRA,Torres2014NJP,Torres2014PRE,Torres2014PRAb,TorresKollmar2015,Torres2014AIP,Torres2016Entropy}.  Fast dynamics, such as Gaussian and exponential decays, emerges in systems with and without correlated eigenvalues. In fact, the same is true for the linear increase of entropies in time~\cite{Santos2012PRL,Santos2012PRE,Torres2016Entropy} and, equivalently, for the  exponential growth of the participation ratio. Therefore,  if exponential behaviors at the quantum level are to be linked with classical chaos, then either level repulsion ceases to be essential for many-body quantum chaos, as it is for one-body quantum chaos, or there is something missing in the discussions of these exponential behaviors.

We show that unambiguous signatures of level repulsion emerge only at long times~\cite{Torres2017,Torres2017PTR,TorresPRBR2018}, after the fast dynamics. Long times are required for the dynamics to be able to resolve the discreteness of the spectrum and detect whether the eigenvalues are or not correlated.  We stress, however, that this chapter does not answer the question of whether there is indeed a one-to-one correspondence between level repulsion and many-body quantum chaos. 

The survival probability presents different behaviors at different time scales~\cite{Torres2015,Torres2016BJP,Tavora2016,Tavora2017,Torres2017,Torres2017PTR,
Torres2017AIP,Torres2018Book,TorresPRBR2018}. To better understand the causes of these behaviors in realistic many-body quantum systems, we analyze the survival probability evolving under full random matrices (FRM). These matrices are not realistic, since they imply the interaction of all particles at the same time, but they allow us to extract analytical expressions, which guide our studies of realistic systems~\cite{TorresPRBR2018}.

For both, FRM and realistic chaotic many-body quantum systems, the survival probability shows four different time scales. The Hamiltonian matrices considered are finite, so the dynamics eventually saturates to a value determined by the level of delocalization of the initial state written in the energy eigenbasis. Before this point, we have:  (i) a very fast initial decay, which may be exponential or Gaussian depending on the strength of the perturbation that takes the system out of equilibrium, (ii) a slower power-law evolution caused by the bounds in the spectrum, (iii) a dip below the saturation point caused by the presence of correlated eigenvalues. This dip, known as correlation hole~\cite{Leviandier1986,Guhr1990,Wilkie1991,Alhassid1992,Gorin2002}, is nonexistent in systems that have uncorrelated eigenvalues or continuous spectrum.

%% Model %%%%%%%%%%%%
\section{Models} 
\label{sec:Models}

We consider systems described by the Hamiltonian 
\begin{equation}
H=H_0 + J V ,
\end{equation}
where $H_0$ is an integrable Hamiltonian corresponding to the unperturbed part of the total Hamiltonian $H$, and $V$ is a perturbation of strength $J$.  We set $\hbar =1$ and $J=1$. 

We denote the eigenvalues and eigenstates of $H$ as $E_{\alpha}$ and $|\alpha\rangle$, respectively. 

\subsection{Full random matrix}
We study the evolution under FRM from a Gaussian orthogonal ensemble (GOE)~\cite{Dyson1962,MehtaBook,Guhr1998}. They are ${\cal D} \times {\cal D} $ real and symmetric matrices with entries from a Gaussian distribution with mean zero. The variance for the diagonal elements, which constitute $H_0$, is 
\[
\langle H_{ii}^2 \rangle =  2
\]
 and the variance of the off-diagonal elements, which compose $V$, is 
\[
\langle H_{ij}^2 \rangle = 1 \hspace{0.3 cm} \text{for} \hspace{0.3 cm}  i\neq j \nonumber 
\]

\subsubsection{Density of States: Semicircle}
\begin{figure}[ht]  
\includegraphics*[scale=.6]{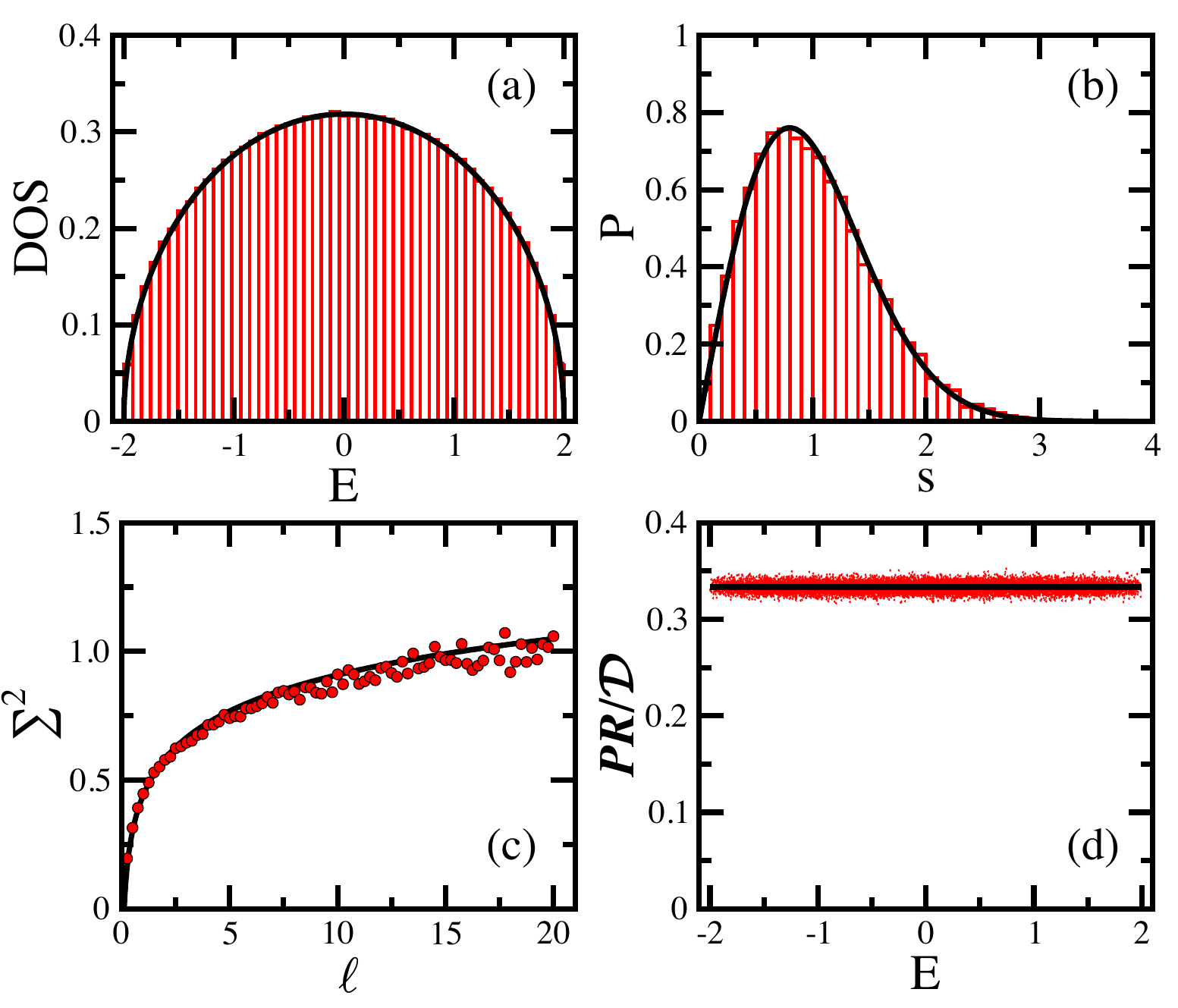}
\caption{Results for a single GOE matrix. (a) Numerical DOS (shade area) compared with Eq. \eqref{DOSfrm} (solid curve).  (b) Numerical $P(s)$ (shaded area) and Eq.\eqref{PsFRM} (solid curve). (c) Level number variance: circles are numerical results and solid curve is Eq.\eqref{SigFRM}. (d) Normalized PR vs $E$: dots are numerical results and the solid line is the analytical result $1/3$.  We divide the eigenvalues by the width of the DOS for a better comparison with the results for the disordered spin model. The dimension of the matrix is ${\cal{D}}=12\,870$.}
\label{Fig:FRM_DOS}   
\end{figure}

The density of states of FRM follows the standard semicircle distribution~\cite{Wigner1955},
\begin{equation}
\text{DOS} (E)= \frac{2}{\pi {\cal E}} \sqrt{1- \left( \frac{E}{{\cal E}} \right)^2 },
\label{DOSfrm}
\end{equation}
where $2 {\cal E}$ is the length of the spectrum, that is $-{\cal E} \leq E \leq {\cal E}$. We show $\text{DOS} (E)$ for one realization of a GOE FRM in Fig.~\ref{Fig:FRM_DOS}  (a)

\subsubsection{Correlatated Eigenvalues}

The eigenvalues of FRM are correlated, prohibited from crossing. This level repulsion is evident in the distribution $P(s)$ of the spacings $s$ between neighboring levels. After unfolding the spectrum~\cite{Gubin2012}, one finds for GOE FRM the following Wigner-Dyson distribution~\cite{Guhr1998},
\begin{equation}
P(s)= \frac{\pi}{2} s\exp \left(- \frac{\pi}{4}s^2 \right).
\label{PsFRM}
\end{equation}
We show $P(s)$ for one realization of a GOE FRM in Fig.~\ref{Fig:FRM_DOS}  (b).

The level spacing distribution $P(s)$ detects short-range correlations between the eigenvalues. There are quantities that capture also long-range correlations, such as the level number variance $\Sigma^2 (\ell) $. Studying different quantities, we get a broader picture of the spectrum. The level number variance is the variance of the number of unfolded eigenvalues in a given interval of length $\ell$. For FRM, $\Sigma^2 (\ell) $ grows logarithmically with $\ell$. For the specific case of GOE FRM, the dependence on $\ell$ is
\begin{equation}
\Sigma^2 (\ell) = \frac{2}{\pi^2} \left[ \ln(2 \pi \ell) + \gamma_e + 1 -\frac{\pi^2}{8} \right] ,
\label{SigFRM}
\end{equation}
where $\gamma_e = 0.5772\ldots $ is Euler's constant. We show $\Sigma^2 (\ell) $ for one realization of a GOE FRM in Fig.~\ref{Fig:FRM_DOS}  (c).

\subsubsection{Delocalized Eigenstates}

The eigenstates $|\alpha \rangle = \sum_n C_n^{\alpha} | n \rangle$ of FRM are random vectors, that is, independently of the chosen basis $|n\rangle $,  the components $C_n^{\alpha} $ are random numbers restricted to the normalization condition, $\sum_n |C_n^{\alpha}|^2 =1$. 
For GOE FRM, the components are real random numbers from a Gaussian distribution. 

To quantify how much delocalized a state is in a certain basis, we use measures such as the participation ratio, defined as
\begin{equation}
PR^{\alpha} = \frac{1}{\sum_{k=1}^{\cal D} |C_k^{\alpha}|^4}.
\end{equation}
For GOE FRM, one can show that $PR = ({\cal D}+2)/3$. In  Fig.~\ref{Fig:FRM_DOS}  (d), we show $PR$ for all eigenstates of one realization of a GOE FRM. As one sees, all eigenstates are equally delocalized, apart from small fluctuations.

\subsection{Disordered spin-1/2 model}
The realistic many-body system that we analyze is a one-dimensional (1D) system of interacting spins-1/2. It has $L$ sites, periodic boundary conditions, and onsite disorder. The unperturbed part of $H$ is
\begin{equation}
H_0= \sum_{k=1}^L h_k  S_k^z   + \sum_{k=1}^L  S_k^z S_{k+1}^z ,
\label{H0_spin}
\end{equation}
where $S_k$ are the spin operators on site $k$.
The Zeeman splittings $h_k$ are random numbers from a uniform distribution $[-h,h]$ and $h$ is the disorder strength. We refer to the eigenstates of $H_0$ as site-basis vectors, also known as computational basis vectors. They correspond to states that on each site has either a spin pointing up in the $z$-direction or pointing down.
The perturbation is given by
\begin{equation}
V=  \sum_{k=1}^L  \left( S_k^x S_{k+1}^x + S_k^y S_{k+1}^y \right) .
\label{ham}
\end{equation}
In the site-basis, $V$ constitues the off-diagonal part of $H$.

The total spin Hamiltonian $H$ conserves the total spin in the $z$-direction, ${\cal S}^z=\sum_kS_k^z$. We study the largest subspace, ${\cal S}^z=0$, which has dimension ${\cal D}=L!/(L/2)!^2$. 

The disordered spin Hamiltonian has been considered in studies of spatial localization since 2004 \cite{Santos2004,SantosEscobar2004,Santos2005loc,Dukesz2009}. An advantage of using it for studies of dynamics is that we deal with averages over disorder realizations, which is a way to reduce finite size effects and smoothen the curves describing the evolution~\cite{Torres2015,Torres2017}. This is particularly important in studies of long-time evolutions, as done here.

\subsubsection{Density of States: Gaussian}

\begin{figure}[ht]
\includegraphics*[scale=.6]{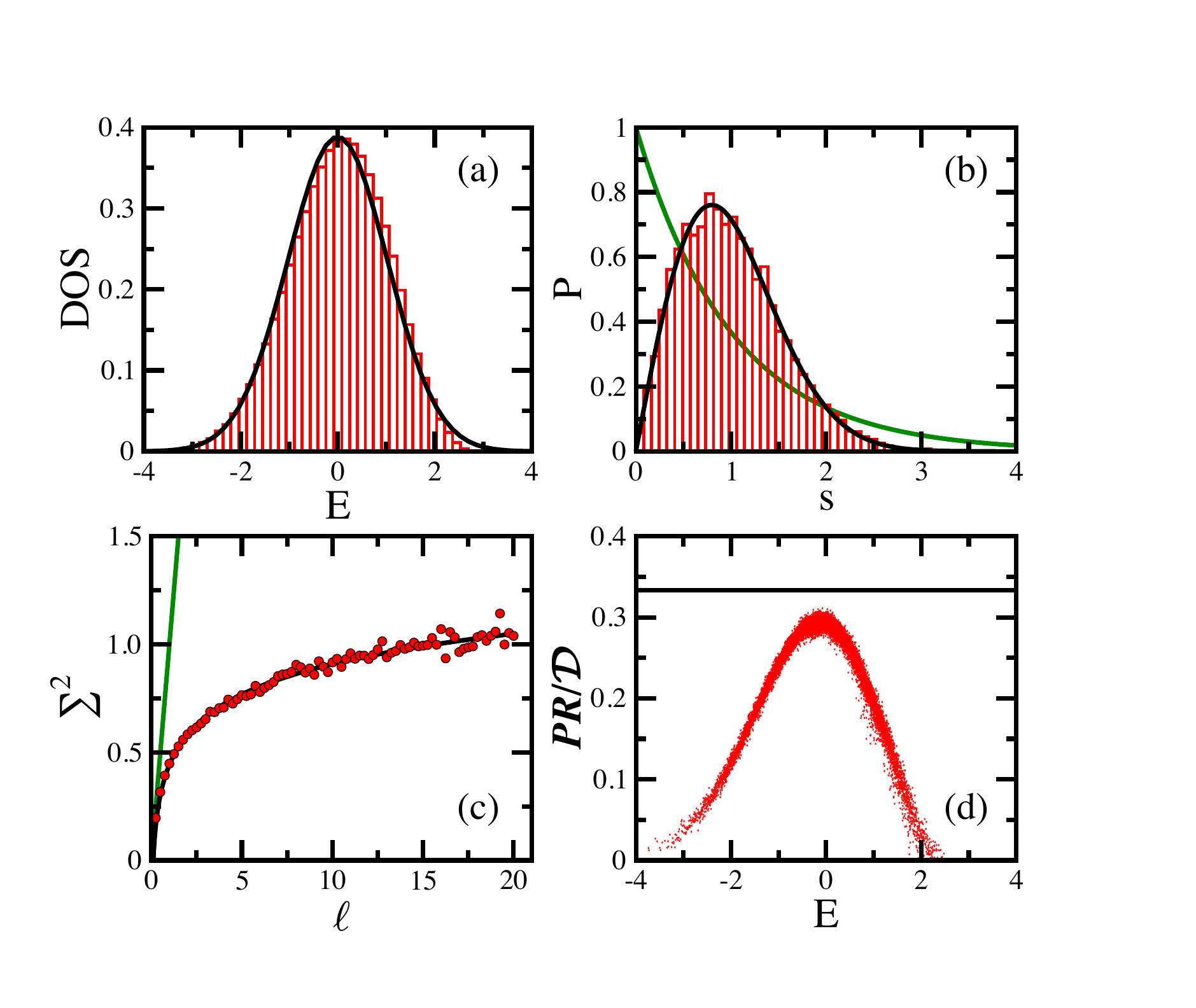}
\caption{Results for a single realization of the realistic disordered spin model with $h=0.5$. (a) Numerical DOS (shaded area) compared with a Gaussian distribution (solid curve).  (b) Numerical $P(s)$ (shaded area), Wigner-Dyson distribution Eq.\eqref{PsFRM} (black solid curve), and Poissonian distribution $P(s)=\exp(-s)$ (green solid curve). (c) Level number variance: circles are numerical results and solid curves are Eq.\eqref{SigFRM} (black) and $\Sigma^2 (\ell)=\ell$ (green). (d) Normalized PR vs $E$: dots are numerical results and the solid line is the analytical result $1/3$ for GOE FRM.  The eigenvalues are rescaled as in Fig.~\ref{Fig:FRM_DOS}, by dividing them by the width of the DOS; ${\cal{D}}=12,870$.}
\label{Fig:spin_DOS}     
\end{figure}

The density of states of delocalized many-body quantum systems with few-body (in our case, two-body) interactions is Gaussian~\cite{Brody1981}. This can be shown analytically in the case of integrable models~\cite{SchiulazARXIV}. In Fig.~\ref{Fig:spin_DOS} (a), we show  $\text{DOS}(E)$ for one realization of the disordered spin model with $h=0.5$.

\subsubsection{Eigenvalues}

The properties of the spectrum of the spin model depend on the strength of the disorder. When $h$ is of the order of $J$, the eigenvalues are correlated and show a Wigner-Dyson distribution. For the system size considered here ($L=16$,  ${\cal D} = 12870$), strong level repulsion occurs for $h=0.5$ \cite{Torres2017}, as depicted in Fig.~\ref{Fig:spin_DOS} (b).

In integrable models, the eigenvalues are uncorrelated and the levels are not prohibited from crossing. Usually, the level spacing distribution  is Poissonian, $P(s)=\exp(-s)$. In the absence of disorder, $h=0$, the eigenvalues are uncorrelated. A Poissonian distribution also emerges for $h>h_c$, where $h_c$ is a critical point above which the system becomes localized in space. Away from the integrable and chaotic points, the level spacing distribution is intermediate between Poissonian and Wigner-Dyson~\cite{Torres2017}.

When it comes to the level number variance, we find that deep in the chaotic regime ($h=0.5$), $\Sigma^2 (\ell)$ agrees well with the results for the GOE FRM for the values of $\ell$ shown in Fig.~\ref{Fig:spin_DOS} (c). For uncorrelated eigenvalues, the level number variance grows linearly with $\ell$, that is $ \Sigma^2 (\ell) =\ell$. In the case of our spin model, as the disorder strength moves away from the chaotic region and approaches the integrable limits ($h>h_c$ or $h\rightarrow 0$), the results for the level number variance get out of the logarithmic curve for smaller values of $\ell$, being followed by a behavior that is closer to linear than logarithmic~\cite{BertrandPRB2016}.

\subsubsection{Eigenstates}

The components of the eigenstates of realistic systems are never completely uncorrelated as in FRM, so we do not have $PR \sim {\cal D}/3$. However, in the chaotic domain, we do find $PR \propto {\cal D}$ \cite{Torres2015,Torres2017}, with pre-factors smaller than 1/3. 

Contrary to FRM, where the basis is ill defined, the values of the $PR$ for realistic systems depend on the chosen basis. The appropriate basis to analyze the transition to chaos is the mean-field basis, as discussed in Refs.~\cite{Santos2012PRL,Santos2012PRE,Borgonovi2016,Torres2016Entropy}. In studies of localization in space, the site-basis vector is a natural choice.

The level of delocalization of the eigenstates of realistic systems also depends on their energy. Eigenstates closer to the middle of the spectrum tend to be more delocalized than those closer to the edges. In Fig.~\ref{Fig:spin_DOS} (d) we show $PR$ for all eigenstates of one realization of the disordered spin model in the chaotic limit, $h=0.5$.

%%%%%%%%%%%%%%%
\section{Survival Probability}
%%%%%%%%%%%%%%%

We prepare the system in an eigenstate $|n_0\rangle \equiv  |\Psi(0) \rangle$ of $H_0$ and let it evolve according to the total Hamiltonian $H$, that is
\[
|\Psi(t) \rangle = e^{-i H t} |\Psi(0) \rangle = \sum_{\alpha} C_{n_0}^{\alpha} e^{-i E_{\alpha} t} |\alpha\rangle,
\]
where  $C^{\alpha}_{n_0}= \langle \psi_{\alpha} |  \Psi(0) \rangle $ is the overlap between the initial state and the energy eigenbasis.

From the many available observables, we study one of the simplest from which a lot of information can be obtained, the survival probability.  Matematically it is given by 
\begin{equation}
W_{n_0} (t) =\left| \langle \Psi(0) |  \Psi(t) \rangle \right|^2= \left| \sum_{\alpha}  \left| C_{n_0}^{\alpha} \right|^2 e^{-i E_{\alpha}t} \right|^2 .
\label{Eq:SP}
\end{equation}
Physically, it is the probability of finding the initial state at time $t$.
The sum in Eq.~(\ref{Eq:SP}) can be written in terms of an integral as
\begin{eqnarray}
W_{n_0}(t) =  \left| \int dE\, e^{ - iEt} \rho_{n_0}(E) \right|^2,
\label{saitorho}
\end{eqnarray} 
where 
\begin{equation}
\rho_{n_0}(E)  \equiv \sum_{\alpha}  | C^{\alpha}_{n_0} |^2 \delta (E - E_\alpha )
\end{equation} 
is the energy distribution weighted by the components $| C^{\alpha}_{n_0} |^2$ of the initial state. It is called local density of states (LDOS). The survival probability is  the absolute square of the Fourier transform of the LDOS.

The energy of the initial state is the mean of the LDOS, 
\begin{equation}
E_{n_0}= \langle \Psi (0)| H |\Psi (0)\rangle = \sum_{\alpha} |C^{\alpha}_{n_0}|^2 E_{\alpha},
\label{energyinitialstate}
\end{equation}
and the variance of the LDOS is
\begin{equation}
\sigma_{n_0}^2=\sum_{\alpha} |C^{\alpha}_{n_0}|^2 (E_{\alpha} - E_{n_0})^2.
\label{stinitialstate}
\end{equation}

At very short times, $t \ll \sigma_{n_0}^{ - 1}$, independently of the model or of the initial state, the survival probability shows a universal quadratic behavior in $t$,
\begin{equation}
W_{n_0}(t)   \approx  1 - \sigma_{n_0}^2 t^2 .
\label{quadractic}
\end{equation}
This is obtained by expanding Eq.~(\ref{Eq:SP}) \cite{Torres2014NJP,Tavora2017}. At later times, the behavior of the survival probability depends on the time scale.

%%%%%%%%%%%%%%%
\section{Dynamics of Chaotic Systems}
%%%%%%%%%%%%%%%

To better understand the different behaviors of the survival probability at different time scales, we compare the results obtained for the evolution under FRM with those for the disordered spin-1/2 model in the chaotic limit, $h=0.5$. 

\subsection{Fast initial decay}

Beyond the universal quadratic decay, the evolution of the survival probability is initially controlled by the shape of the envelope of the LDOS.

\subsubsection{Full random matrices}
\begin{figure}[ht]
\includegraphics*[scale=.6]{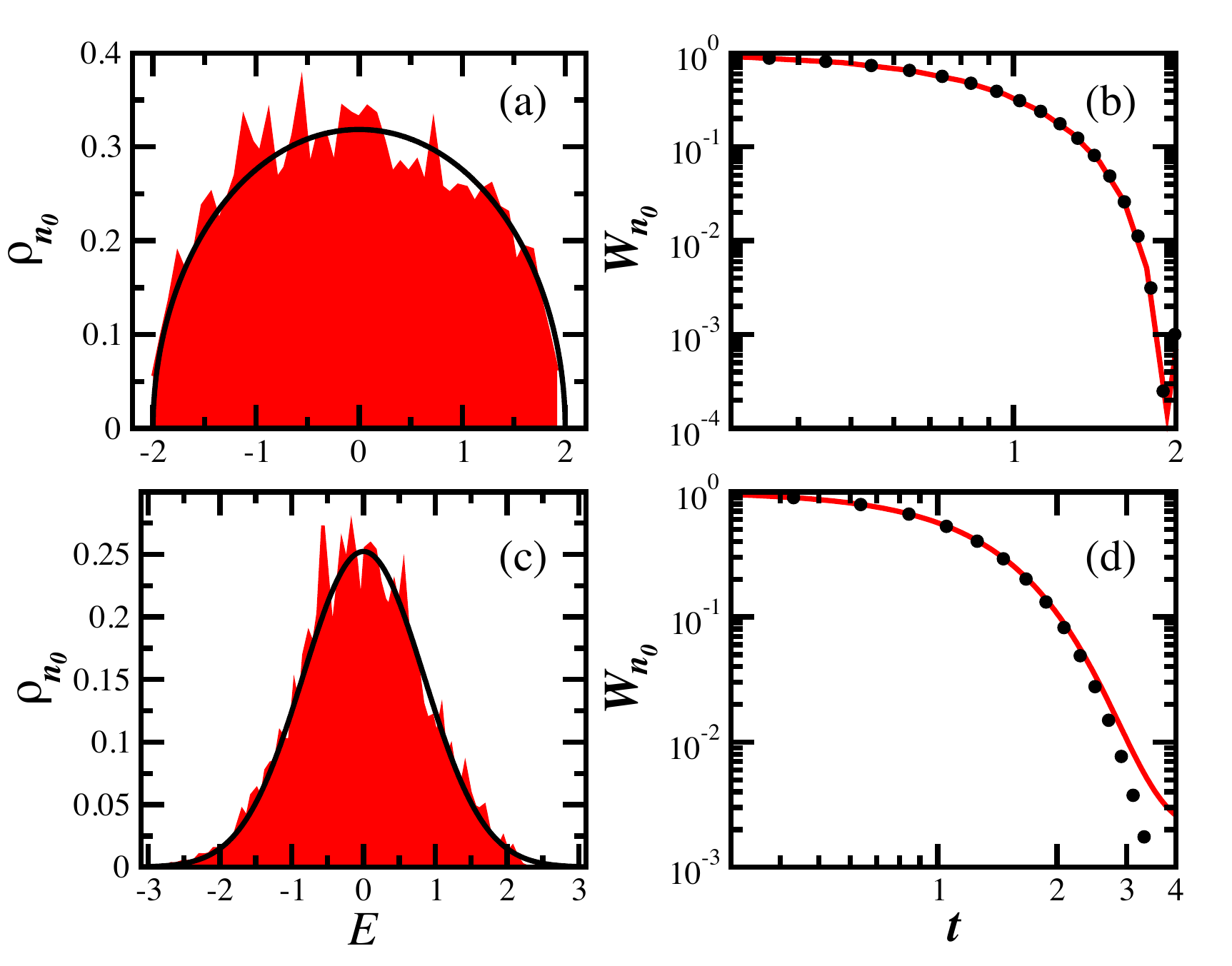}
\caption{Local density of states (a,c) and survival probability (b,d) for the  GOE FRM (top) and the disordered spin model (bottom) with $h=0.5$. For both, the initial state $|\Psi(0)\rangle$ has energy $E_{n_0}$ close to zero. In (a,c) the red shaded areas are numerical results for one realization and the black solid lines are the semicircle (a) and the Gaussian function (c).  In (b,d) the red solid curves are numerical results and the dots correspond  to Eq.\eqref{Eq:semi} in (b) and Eq.\eqref{eq:GausD}. In (b) an average over $10^4$ data was carried out, while in (d) the average is over $10^5$ data. For all panels ${\cal{D}}=12\,870$.} 
\label{Fig:shortTime}     
\end{figure}

The shape of the LDOS of an arbitrary initial state projected into the eigenstates of a FRM coincides with the shape of the DOS, that is, the envelope is semicircular~\cite{Torres2014PRA,Torres2014NJP,Torres2014PRAb}, 
\begin{equation}\label{eq:LDOS}
\rho_{n_0}(E) =  \frac{1}{\pi \sigma_{n_0}} \sqrt{1 - \left(\frac{E}{2  \sigma_{n_0}}\right)^2}\,,
\end{equation}
where, according to Eq.~(\ref{DOSfrm}),  $4 \sigma_{n_0} = 2 {\cal E} $ is the length of the spectrum. The LDOS is shown in Fig.~\ref{Fig:shortTime} (a).

The Fourier transform of a semicircle gives~\cite{Torres2014PRA,Torres2014NJP,Torres2014PRAb} 
\begin{equation}
W_{n_0}(t) = \frac{ {\cal J}_1(2 \sigma_{n_0} t)^2}{(\sigma_{n_0} t)^2},
\label{Eq:semi}
\end{equation} 
where ${\cal J}_1(t)$ is the Bessel function of first kind. This leads to a very fast initial decay, as shown in Fig.~\ref{Fig:shortTime} (b). This is the fastest decay we find for many-body quantum systems with a discrete spectrum and a single peak LDOS. Faster decays can be obtained with LDOS that have more than one peak~\cite{Torres2014PRAb}.

\subsubsection{Chaotic spin-1/2 model}

For the chaotic spin model, we select as initial states, eigenstates from $H_0$ (\ref{H0_spin}) with energy $E_{n_0}$ very close to the middle of the spectrum. The LDOS for these states, just as the DOS, has a Gaussian shape, as shown in Fig.~\ref{Fig:shortTime} (c).

The Fourier transform of a Gaussian gives a Gaussian decay, 
\begin{equation}
W_{n_0}(t) = \exp(-\sigma_{n_0}^2 t^2), 
\label{eq:GausD}
\end{equation}
as indeed seen in Fig.~\ref{Fig:shortTime} (d). The curve is an average over several initial states and disorder realizations that total $10^5$ data. We stress that the decay is truly Gaussian and not only the quadratic behavior at short times.

\begin{figure}[ht]
\includegraphics*[scale=.6]{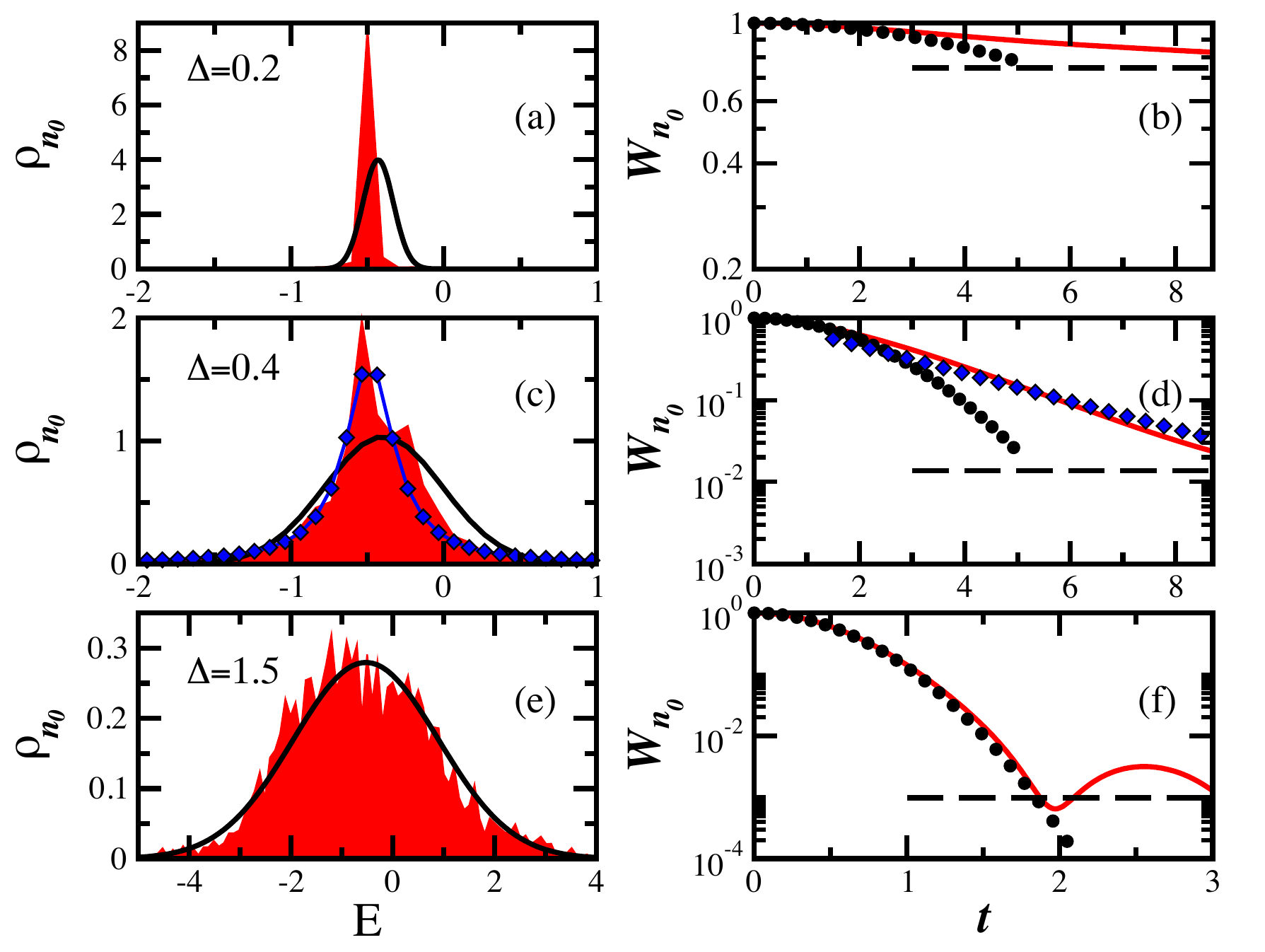}
\caption{Local density of states (left column) and survival probability (right column) for the  XXZ model [Eq.~\eqref{Hxxz}]. The values of $\Delta$ are indicated in the figure. The initial state $|\Psi(0)\rangle$ has energy far from the edge of the spectrum and it is an eigenstate of the XX  model [Eq.~\eqref{eq:XX}]. Black solid line (left) and black circles (right): Gaussian LDOS and Gaussian decay, respectively with $\sigma_0$ from Eq.~\eqref{stinitialstate}. Red shaded area (left) and red solid lines (right): numerical results. Blue diamonds in (c) and (d): Lorentzian fit and exponential decay, respectively. $L=18$, $N_\text{up}=6$, open chain. Horizontal dashed lines indicate the saturation value [see Eq.~\eqref{eq:sat}].}
\label{Fig:integrable_chaos}     
\end{figure}

The decay functions for FRM and realistic models are, of course, different, but the cause of the behaviors are the same, namely the shape of the envelope of the LDOS. We note that the shape depends on the model, on the energy of the initial state, and on the strength of the perturbation, but not on the existence of level repulsion~\cite{Torres2014PRA,Torres2014NJP,Torres2014PRE,Torres2014PRAb}. Integrable realistic models can also lead to Gaussian or exponential decays. To illustrate this fact, we show in Fig.~\ref{Fig:integrable_chaos}, the survival probability for the integrable XXZ model with open boundary conditions described by
\begin{equation}
H_{XXZ}=  \sum_{k=1}^{L-1}  \left( S_k^x S_{k+1}^x + S_k^y S_{k+1}^y + \Delta S_k^z S_{k+1}^z \right)   .
\label{Hxxz}
\end{equation}
The anisotropy parameter $\Delta $ is the strength of the perturbation.  We take as initial state one with $E_{n_0} \sim 0$ from the eigenstates of the XX model given by
\begin{equation}
\sum_{k=1}^{L-1}  \left( S_k^x S_{k+1}^x + S_k^y S_{k+1}^y \right) . 
\label{eq:XX} 
\end{equation} 
When $\Delta$ is very small, the initial state is not so different from one of the energy eigenbasis and the LDOS is very narrow [Fig.~\ref{Fig:integrable_chaos} (a)], resulting in a very slow evolution [Fig.~\ref{Fig:integrable_chaos} (b)]. As $\Delta$ increases, the LDOS broadens and becomes Lorentzian [Fig.~\ref{Fig:integrable_chaos} (c)], which leads to the exponential behavior of the survival probability [Fig.~\ref{Fig:integrable_chaos} (d)]. If we increase $\Delta$ even further, the LDOS broadens even more, eventually approaching the Gaussian shape of the DOS [Fig.~\ref{Fig:integrable_chaos} (e)]. This is the limit of very strong perturbation, which causes the Gaussian decay [Fig.~\ref{Fig:integrable_chaos} (f)]. In Fig.~\ref{Fig:integrable_chaos} we choose $L=18$ and $N_\text{up} = 6$ spins pointing up (1/3 filling), leading to the total dimension ${\cal{D}}=18\,564$.

 \subsection{Power-law decays at intermediate times}
 
Any quantum system has at least one bound in the spectrum corresponding to the ground state. In the case of finite systems, there are two energy bounds. The presence of these bounds cause the eventual partial reconstruction of the initial state, which slows down the evolution of the survival probability. We then find power-law decays~\cite{MugaBook} even in the most chaotic scenario of FRM. The value of the power-law exponent depends on how the LDOS approaches the energy bounds.

If the LDOS goes to zero at the bound $E_{low}$ as
\begin{equation}
\rho_{n_0}(E) = (E-E_{low})^{\xi} \eta(E) ,
\end{equation}
with
\[
\lim _{E\rightarrow E_{low}} \eta(E)>0 , 
\]
then the decay of the survival probability at long times is given by
\begin{equation}
W_{n_0}(t) \propto t^{-2(\xi+1)}.
\label{eq:xi_gamma}
\end{equation}
More details can be found in ~\cite{Erdelyi1956,Urbanowski2009,Tavora2016,Tavora2017}.
 
 \subsubsection{Full random matrices}
\begin{figure}[ht]
\includegraphics*[scale=.65]{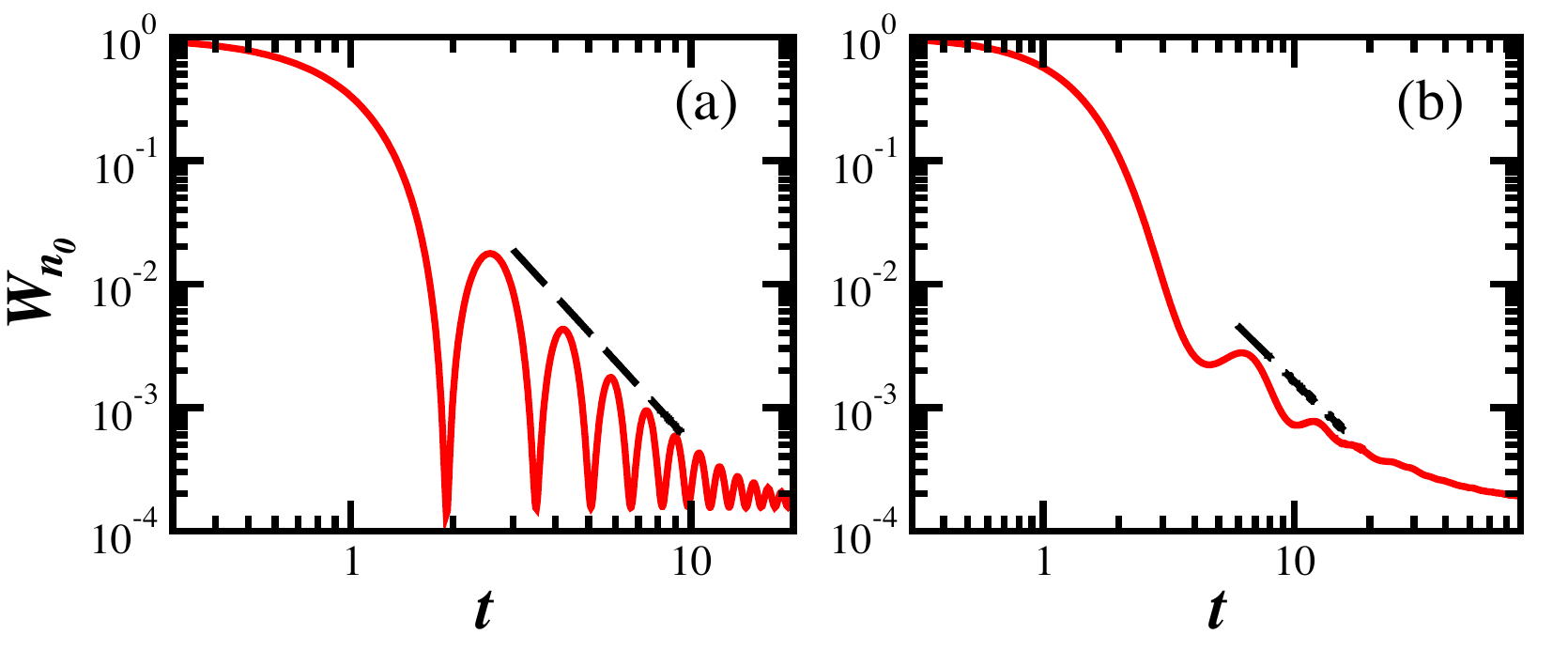}
\caption{Power-law decay of the survival probability. In (a) the GOE FRM: numerical results (solid line) and $t^{-3}$ decay (dashed line). In (b) the disordered spin model with $h=0.5$: numerical results (solid line) and $t^{-2}$ decay (dashed line). The spectrum is rescaled as described in Fig.~\eqref{Fig:FRM_DOS} and Fig.~\eqref{Fig:spin_DOS}. In (a) the average is over $10^4$ data while in (b) the average is over $10^5$ data. For both models ${\cal D}=12\,870$.}
\label{Fig:power}     
\end{figure}

The tails of the LDOS of FRM falls with the square root of the energy, $E^{1/2}$ , so one has $\xi  = 1/ 2$. Following Eq.~\eqref{eq:xi_gamma}, the decay should be $t^{-3}$, as indeed confirmed by Fig.~\ref{Fig:power} (a).  The power-law decay is the envelope of the decay of the Bessel oscillations. 

The $t^{-3}$ behavior can also be directly derived by studying Eq.~\eqref{Eq:semi} asymptotically, $t \gg \sigma_{n_0}^{ - 1}$, as shown in~\cite{TorresKollmar2015,Tavora2016,Tavora2017}. In this case, we find that 
 \begin{equation}
W_{n_0}(t) \to \frac{1 - \sin (4 \sigma_{n_0} t)}{2\pi \sigma_{n_0}^3 t^3},
\end{equation}
which makes evident the power-law decay.

\subsubsection{Chaotic spin-1/2 model}

We find a power-law decay for the FRM, so we should also see it for our realistic chaotic spin model. The source of the algebraic decay is the same, namely the bounds in the spectrum. In chaotic systems,  the LDOS is ergodically filled, so this gives enough time for the dynamics to detect the bounds in the spectrum, before resolving the discreteness of the spectrum.

The LDOS of the chaotic spin model is Gaussian, so the tails become approximately constant in energy. Following Eq.~(\ref{eq:xi_gamma}), one has $\xi =0$, so $W_{n_0}(t) \propto t^{-2}$. This power-law exponent can also be obtained from the direct Fourier transform of the Gaussian LDOS, taking the bounds into account when doing the integral~\cite{Tavora2016,Tavora2017}.

In Fig.~\ref{Fig:power} (b), we show with a dashed line the $t^{-2}$ decay. Similarly to the case of FRM, we also see oscillations decaying with a power-law envelope, but larger systems sizes are needed to make this behavior more obvious. We note that even for FRM, as the system size decreases, the oscillations also become less evident~\cite{CampoPRD2017}.

 \subsection{Correlation hole}
 
 We deal with finite systems, so the evolution of the survival probability eventually saturates to its infinite-time average,
 \begin{equation}
 \overline{W}_{n_0} = \sum_{\alpha} \left| C_{n_0}^{\alpha} \right|^4,
 \label{eq:sat}
 \end{equation}
 which is the inverse of the participation ratio for the initial state written in the basis of the energy eigenstates. For FRM,  $\overline{W}_{n_0} = 3/({\cal D}+2)$ and larger values are reached for realistic models.
 
But before saturating, the dynamics resolves the discreteness of the spectrum. In the case of chaotic systems, the correlations between the eigenvalues are reflected in the evolution of the survival probability in the form of a dip below the saturation point, known as correlation hole~\cite{Leviandier1986,Pique1987,Guhr1990,Hartmann1991,Delon1991,Wilkie1991,Alhassid1992,Lombardi1993,Kudrolli1994,Alt1997,Michaille1999,Gorin2004,Alhassid2006,Leyvraz2013}. 
The correlation hole was studied in the context of FRM, billiards, and molecules. We have extended these studies to lattice spin-1/2 models~\cite{Torres2017,Torres2017PTR,Torres2017AIP,Torres2018Book,TorresPRBR2018}.
 
\subsubsection{Full random matrices}

To understand the origins of the correlation hole, one needs to write Eq.~(\ref{Eq:SP}) as
\begin{equation}
W_{n_0} (t) =  \sum _{\alpha_1,\alpha_2} |C^{\alpha_1}_{n_0} |^2 | C^{\alpha_2}_{n_0} |^2 e^{i E_{\alpha_1} t} e^{-i E_{\alpha_2} t} 
= \int  G(E) e^{-i E t} dE + \overline{W}_{n_0} ,
\label{FourierG}
\end{equation}
where
\begin{equation}
G(E)= \sum _{\alpha_1 \neq \alpha_2} |C^{\alpha_1}_{n_0} |^2 | C^{\alpha_2}_{n_0} |^2 \delta( E - E_{\alpha_1} + E_{\alpha_2}  ) 
\label{Eq:GE}
\end{equation}
is the spectral autocorrelation function.  
  
When dealing with FRM, we usually perform averages $\langle . \rangle_{\rm{FRM}} $ over ensembles of random matrices. 
Since the eigenvalues and eigenstates of FRM are statistically independent, we can write $G(E)$  as
\begin{equation}
\langle G(E) \rangle_{\rm{FRM}} = \left\langle \sum _{\alpha_1 \neq \alpha_2} |C^{\alpha_1}_{n_0} |^2 | C^{\alpha_2}_{n_0} |^2 \right\rangle_{\rm{FRM}}\langle \delta( E - E_{\alpha_1} + E_{\alpha_2}  ) \big\rangle_{\rm{FRM}} . 
\label{eq:G}
\end{equation}
The first factor only depends on the random real components of the initial state,
\begin{equation}
\left\langle \sum _{\alpha_1 \neq \alpha_2} |C^{\alpha_1}_{n_0} |^2 | C^{\alpha_2}_{n_0} |^2 \right\rangle_{\rm{FRM}} \!\!\!=  1 - \left\langle \sum_{\alpha} |C_{n_0}^{\alpha} |^4 \right\rangle_{\rm{FRM}} 
= 1- \left\langle \overline{W}_{n_0} \right\rangle_{\rm{FRM}} .
\label{eqCC}
\end{equation} 
The second factor in  Eq.~\eqref{eq:G} is the one that captures the correlations between the eigenvalues. It can be written as
\begin{equation}
\langle \delta( E - E_{\alpha_1} + E_{\alpha_2}  ) \rangle_{\rm{FRM}} =\dfrac{{\cal D}!}{({\cal D}-2)!} \int \int \delta(E - E_{\alpha_1} + E_{\alpha_2}) R_2(E_{\alpha_1},E_{\alpha_2}) dE_{\alpha_1} dE_{\alpha_2},
\end{equation} 
where $R_2(E_{\alpha_1},E_{\alpha_2})$ is the Dyson two-point correlation function. It gives the probability density of finding an energy level around each of the energies $E_{\alpha_1}$, $E_{\alpha_2}$.  The two-point correlation function can be divided as
\begin{equation}
R_2(E_{\alpha_1},E_{\alpha_2}) = R_1(E_{\alpha_1}) R_1(E_{\alpha_2}) - T_2(E_{\alpha_1} , E_{\alpha_2}),
\end{equation}
where $R_1$ is the DOS and $T_2(E_{\alpha_1} , E_{\alpha_2})$ is the two-level cluster function~\cite{MehtaBook}.

The Fourier transform of $R_1$ leads to the behavior described by Eq.~(\ref{Eq:semi}). The Fourier transform of $T_2$ gives the two-level form factor, which for GOE matrices reads
\begin{equation}
b_2(\overline{t}) = [1-2\overline{t} + \overline{t} \ln(1+2 \overline{t})] \Theta (1- \overline{t}) 
+ \{-1 + \overline{t} \ln [ (2 \overline{t}+1)/(2 \overline{t} -1) ] \} \Theta(\overline{t}-1),
\end{equation}
where $\Theta$ is the Heaviside step function  (see details of the derivation in \cite{MehtaBook,Alhassid1992,Torres2017AIP}). The $b_2$ function is responsible for the correlation hole. The Bessel decay in Eq.~(\ref{Eq:semi}) is interrupted by $b_2$, which, after the hole, brings the survival probability to the saturation point.

The two-level form factor $b_2(\overline{t})$ differs from zero only if the eigenvalues have some degree of correlation. It is therefore an unambiguous signature of the presence of level repulsion that one finds by studying the time evolution of the system~\cite{Torres2017PTR}.  If the eigenvalues are uncorrelated,  $b_2(\overline{t})=0$.

\begin{figure}[ht!]
\includegraphics*[scale=.65]{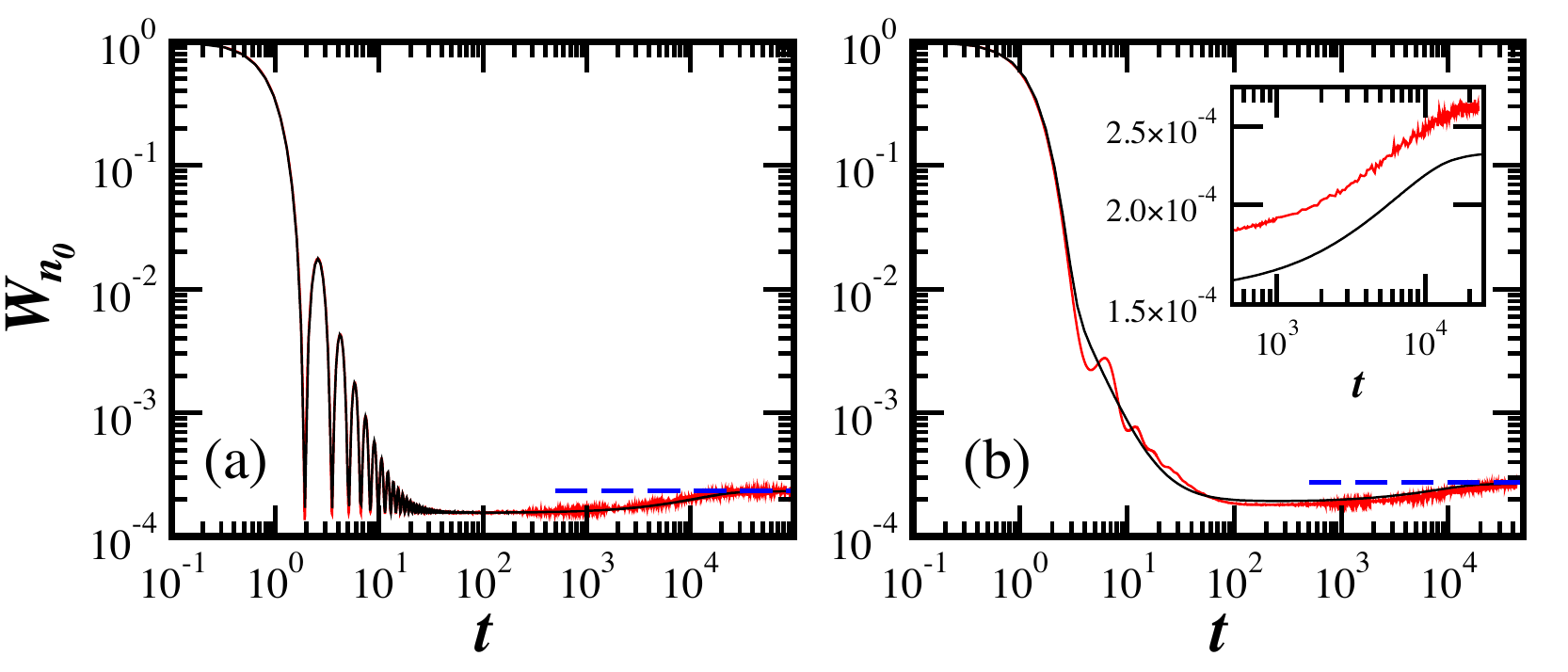}
\caption{Survival probability. (a) Numerical results for GOE FRM (red) and Eq.~(\ref{Eq:Wo}) (black) are superposed. (b)  Numerical results for the disordered model with $h=0.5$ (red) and the fitting curve Eq.\eqref{Eq:Wspin} (black). In  both panels the blue dashed lines are the corresponding saturation values. The spectrum is rescaled as described in Fig.~\eqref{Fig:FRM_DOS} and Fig.~\eqref{Fig:spin_DOS}. Inset of (b) shows the growth of $W_{n_0}(t)$ from the bottom of the correlation hole to saturation: Eq.~(\ref{Eq:Wo}) (bottom black curve) and numerical results for $h=0.5$ averaged over small time windows (top red curve). In (a) average over $10^4$ disorder realizations. In (b) the average is over $1\,287$ different initial states and $77$ different disorder realizations, adding up to $10^5$ data. For both panels ${\cal D}=12\,870$.}
\label{Fig:hole}     
\end{figure}

The complete analytical expression for the survival probability evolving under GOE FRM of large dimensions was shown in Ref.~\cite{TorresPRBR2018}. We have
\begin{equation}
\langle W_{n_0} (t) \rangle_{\rm{FRM}}  =  \frac{1-\langle \overline{W}_{n_0} \rangle_{\rm{FRM}} }{{\cal D} -1} \left[ 4 {\cal D} \frac{{\cal J}_1^2 ({\cal E} t)}{({\cal E} t)^2} 
-  b_2 \left( \frac{{\cal E} t}{4 {\cal D} } \right) \right] + \langle \overline{W}_{n_0} \rangle_{\rm{FRM}} .
\label{Eq:Wo}
\end{equation}
This expression matches the numerical result extremely well, as seen in Fig.~\ref{Fig:hole} (a).

\subsubsection{Chaotic spin-1/2 model}

The correlation hole is also evident in the chaotic spin model. As seen in the inset of Fig.~\ref{Fig:hole} (b), the curve for the spin model is parallel to the analytical one for FRM in Eq.~(\ref{Eq:Wo}). 
Note that the spectra of the FRM and for the spin model are rescaled for a better comparison. This implies that the same $b_2$ function can be used to describe the long-time behavior of the survival probability of realistic chaotic models. At such long times, the dynamics no longer depends on details, such as shape of the DOS and LDOS and the structure of the eigenstates, but cares only about the correlations between the eigenvalues.

Motivated by the comparisons between our results for the FRM and the chaotic spin model, we propose an expression to describe the entire evolution of the survival probability for the spin model. It is written as
\begin{equation} 
\langle W_{n_0} (t) \rangle =  \frac{1- \langle \overline{W}_{n_0} \rangle}{{\cal D} -1} \left[ {\cal D} \frac{g(t)}{g(0)} 
- b_2 \left(\frac{\sigma_{n_0} t}{ {\cal D}}\right) \right] + \langle \overline{W}_{n_0} \rangle ,
\label{Eq:Wspin}
\end{equation}
where 
\begin{equation}
g(t) = e^{-\sigma_{n_0}^2 t^2} + A \frac{1-e^{-\sigma_{n_0}^2 t^2} }{\sigma_{n_0}^2 t^2}
\end{equation}
 and $A$ is a fitting constant. The function $g(t)$ is used to capture the initial Gaussian decay and the subsequent power-law behavior $\propto t^{-2}$. The curve from Eq.~\eqref{Eq:Wspin} in shown in Fig.~\ref{Fig:hole} (b) together with the numerical result. It is quite impressive that we can match the entire evolution so well using  a single fitting constant.

%%%%%%%%%%%%%%%
\section{Dynamics: From Chaos to Localization}
%%%%%%%%%%%%%%%

\begin{figure}[ht]
\includegraphics*[scale=.65]{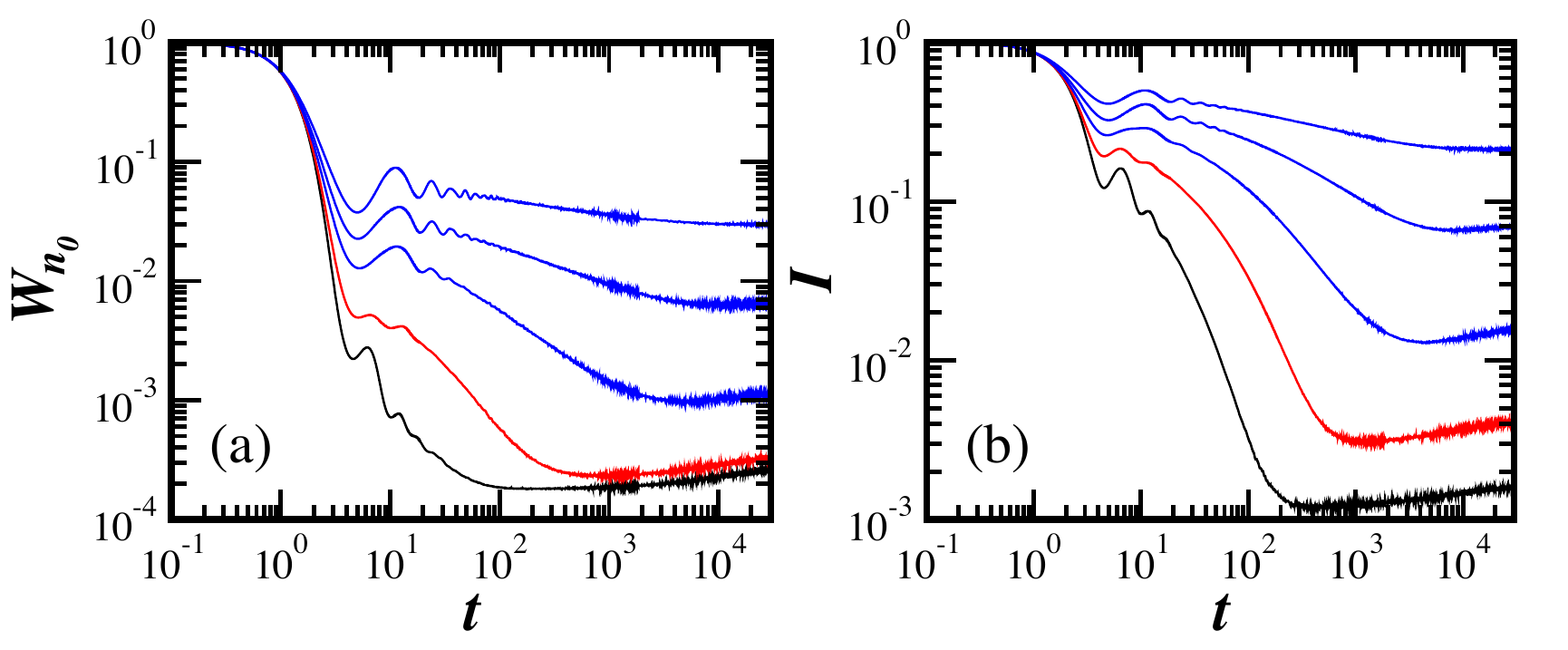}
\caption{Full decay of the survival probability (a) and of the spin density imbalance (b) for the disordered spin model. The numerical results from bottom to top have $h=0.5, 1, 1.5, 2, 2.5$. In (a), the average is over $10^5$ data as in Fig.~\eqref{Fig:hole} (b). In (b) the average is over $10^4$ data. For both panels ${\cal D}=12\,870$. }
\label{Fig:disorder}     
\end{figure}

As we increase $h$ above 0.5 and the system approaches the localized phase in space, the evolution of the survival probability slows down and saturates at a higher plateau, as shown in Fig.~\ref{Fig:disorder}  (a).

For $0.5\leq h \leq 1$, where the system is chaotic, the power-law exponent $\gamma$ of $W_{n_0} (t) \propto t^{-\gamma}$ is $2\leq \gamma\leq 1$. As we said above, $\gamma =2$ is caused by the energy bounds, the values $2< \gamma\leq 1$ must be caused by a combination of bound effects and possible small correlations between the eigenstates~\cite{Tavora2016,Tavora2017}.

When $h>1$, the LDOS becomes sparse and the initial states are no longer fully delocalized in the energy eigenbasis, that is $PR_{n_0} = \overline{W}_{n_0}^{-1}\propto {\cal D}^{D_2}$, where $D_2<1$ is the fractal dimension. As shown in Ref.~\cite{Torres2015}, $D_2$ coincides with $\gamma$. This implies that the components $|C_{n_0}^{\alpha}|^2$ are correlated, so that the spectral autocorrelation function in Eq.~(\ref{Eq:GE}) becomes
\begin{equation}
G(E) \propto E^{D_2 -1}.
\end{equation}
The power-law exponent $\gamma = D_2<1$ is a consequence of the fact that the eigenstates become fractal and therefore correlated.

As we mentioned when describing Fig.~\ref{Fig:spin_DOS}, as $h$ increases above 0.5, the level number variance escapes the logarithmic behavior for smaller values of $\ell$ and the level spacing distribution changes gradually from Wigner-Dyson to Poissonian. This occurs because the eigenvalues are becoming less correlated, which is reflected also in the correlation hole. It gets less deep as the system moves away from the chaotic region and the point where the hole starts happens later in time. This point corresponds to the first crossing of $W_{n_0} (t)$ with the saturation line. The gradual shift of the correlation hole to later times is directly related with the earlier point where $\Sigma^2(\ell)$ leaves the logarithmic curve and should be related with the Thouless energy.

We note that the correlation hole is not exclusive to the survival probability. It is found also in other observables, such as the imbalance of the spin density for all sites~\cite{Luitz2016,Lee2017}, 
\begin{equation}
I(t) = \frac{4}{L} \sum_{k=1}^L \langle \Psi(0)| S^z_k(0) S^z_k(t) |  \Psi(0) \rangle .
\label{Eq:Imba} 
\end{equation}
We show results for $I(t)$ in Fig.~\ref{Fig:disorder}  (b) for different values of $h$. The curves are comparable to those seen for the survival probability in Fig.~\ref{Fig:disorder}  (a). The correlation hole is deep and broad in the chaotic limit, $h=0.5$. It shrinks and becomes shallower as $h$ increases~\cite{TorresPRBR2018}. We should expect similar behaviors for other observables, such as the OTOC~\cite{TorresPRBR2018}. In Ref.~\cite{Torres2017PTR} we also saw a sign of the correlation hole in the evolution of the Shannon (information) entropy. There, it appeared as a bulge above the saturation point. It was, however, very small. To observe the correlation hole experimentally, the evolution needs to remain coherent for long times and averages need to be carried out.

%%%%%%%%%%%%%%%
\section{Conclusion}
%%%%%%%%%%%%%%%

Along these past four years, we have developed a detailed picture of the entire evolution of the survival probability for realistic lattice many-body systems. We are now extending this understanding to other observables and hope to be able to suggest equations similar to Eq.~(\ref{Eq:Wspin}) for generic observables. We believe that comparisons with results for FRM, as done here, will be helpful in this pursuit. Despite being unrealistic, FRM allows for analytical results, which can guide us when dealing with realistic models. 

We stress that signatures of level repulsion show up at long times, when the dynamics resolves the discreteness of the spectrum. It remains to show whether level repulsion is essential for many-body quantum chaos, as it is for one-body quantum chaos.  This is an important question for studies of thermalization, since for many years, chaos has been seen as the main mechanism for the relaxation of isolated many-body quantum systems to thermal equilibrium.

\bigskip

\acknowledgments

L.F.S. was supported by the NSF grant No. DMR-1603418. E.J.T.-H. acknowledges funding from CONACyT and VIEP-BUAP, Mexico.  He is also grateful to LNS-BUAP for allowing use of their supercomputing facility.

\end{document}